\documentclass[pdflatex,sn-mathphys-num]{sn-jnl}% Math and Physical Sciences Numbered Reference Style 
\usepackage{graphicx}%
\usepackage{multirow}%
\usepackage{amsmath,amssymb,amsfonts}%
\usepackage{amsthm}%
\usepackage{mathrsfs}%
\usepackage[title]{appendix}%
\usepackage[table]{xcolor}%
\usepackage{bm}
\usepackage{bbm}
\usepackage{nicefrac}
\usepackage{siunitx}
\usepackage{tikz}
\usepackage{enumitem}
\usepackage{comment}
\usepackage{array,multirow}
\usepackage[normalem]{ulem}
\usepackage{setspace}
\usepackage{dcolumn}
\usepackage{lipsum}
\usepackage[numbers,sort&compress]{natbib}
\usepackage{hyperref}
\usepackage{soul}
\usepackage{lineno}
\theoremstyle{thmstyleone}%
\theoremstyle{thmstyletwo}%

\theoremstyle{thmstylethree}%

\begin{document}

\title{
  Inverse-scattering separable \textsl{NN} potential 
  constrained to phase-shift data up to 2.5~GeV.
  I.- Uncoupled states
}
% \title[Article Title]{Article Title}

%\author[1,2]{ H. F. Arellano}
%\author[1]{N. A. Adriazola}
%\affiliation[1]{
%  Department of Physics - FCFM, University of Chile,
%  Av. Blanco Encalada 2008, Santiago, RM 8370449, Chile
%}
%  \affiliation[2]{CEA, DAM, DIF, F-91297 Arpajon, France}

\author*[1,2]{\fnm{H. F.} \sur{Arellano}}\email{arellano@dfi.uchile.cl}

\author[1]{\fnm{N. A.} \sur{Adriazola}} %\email{iiauthor@gmail.com}

\affil*[1]{\orgdiv{Department of Physics - FCFM}, 
           \orgname{University of Chile}, 
           \orgaddress{\street{Av. Blanco Encalada 2008}, 
           \city{Santiago}, 
           \postcode{RM 8370449}, %\state{State}, 
           \country{Chile}}}

\affil[2]{\orgdiv{CEA, DAM, DIF, 
%\orgname{Organization}, 
%       \orgaddress{\street{Street}, \city{City}, 
\postcode{F-91297 Arpajon},
%\state{State}, 
\country{France}}}


%
% This is a template file for The European Physical Journal
%
% Copy it to a new file with a new name and use it as the basis
% for your article
%
%%%%%%%%%%%%%%%%%%%%%%%% Springer-Verlag %%%%%%%%%%%%%%%%%%%%%%%%%%
%
%\usepackage{latexsym}
%
%
%\ead{arellano@dfi.uchile.cl}
\date{Received: date / Revised version: date}
%\linenumbers
\abstract{
  We introduce a new method to construct, within inverse-scattering
  theory, an energy-independent separable potential capable
  of reproducing exactly both phase shift and absorption over a
  predefined energy range.
  The approach relies on the construction of non-overlapping
  multi-rank separable potentials, whose form factors are obtained
  by solving linear equations on intervals where the
  $K$ matrix does have zeros.
  Applications are made to nucleon-nucleon \textit{(NN)} 
  interactions constrained to the {\footnotesize SAID-SP07}
  phase-shift analysis up to 2.5~GeV lab energy.
  The inversion potentials are channel dependent with
  rank dictated by the number of zeros of the $K$ matrix, 
  reproducing the data up to a selected upper momentum.
  The account for absorption yields complex separable
  form factors, resulting in a non-Hermitian potential.
  Applications are restricted to \textit{NN} spin-uncoupled states
  considering a Schr\"odinger-like wave equation with minimal relativity.
  Its extension to spin-coupled states and relativistic kernels
  are discussed.
% \PACS{
%{02.30.Zz}{Inverse problems} \and
%{13.75.Cs, 13.85.-t}{Nucleon-nucleon interactions}\and
%{11.55.-m}{Scattering matrix}\and
%{03.65.Nk}{Scattering theory (quantum mechanics)}\and
%{12.39.Pn}{Potential models, 12.39.Pn}
%     } % end of PACS codes
}
%\authorrunning{Arellano-Adriazola}
%\titlerunning{ Inverse-scattering separable \textit{NN} potential ...}
\maketitle
%
% \sloppy

% \baselineskip 16pt
% \baselineskip 20pt
% \baselineskip 24pt

\section{Introduction}
The concept of two-body potentials to represent the interaction between
nucleons has proven to be a powerful tool for \textit{ab-initio}
and microscopic studies of nuclear many-body phenomena such as 
nuclear reactions, nuclear structure and nuclear matter equation-of-state.
Soon after Hideki Yukawa introduced his one-pion-exchange 
theory~\cite{Yukawa1935} to describe the nucleon-nucleon (\textit{NN})
interaction,
sustained progress has been made over the years
towards building meson-exchange potential 
models~\cite{Wiringa1995,Lacombe1980,Stocks1994}, 
and more recently, chiral effective field theory interactions
with pionic degrees of 
freedom~\cite{Weinberg1990,Epelbaum2015,Entem2003,Holt2010,Hebeler2010a,Piarulli2015,Entem2020}.

In all these constructions a limited set of parameters are 
adjusted to provide optimal fits to \textit{NN} scattering data 
and static properties of the deuteron,
the only two-nucleon bound state in free space.
A common feature of these realistic potential models is that
they are designed to account for phase-shift data at nucleon 
beam energies of up to about 300~MeV. 
In the context of chiral perturbation theory, the most advanced
developments go up to order of six, becoming limited to lab energies
of 0.5~GeV at the most~\cite{Entem2020,Eyser2004}, where inelasticities
are small.
Thus, the resulting potentials turn out to be Hermitian, consistent with 
the fact that no loss of flux takes place in \textsl{NN} collisions
at these energies.
At higher energies these realistic potentials become unsuitable
as their implied scattering amplitude constitute extrapolations
of the model,
together with their inability to account for the loss of flux 
above the pion production threshold.
From a phenomenological stand point,
such limitations in realistic potentials can be corrected 
by adding an energy- and state-dependent separable 
term with predefined form factors~\cite{Funk2001,Arellano2002}.

In this paper we introduce a novel framework to obtain a separable
representation for the \textit{NN} interaction constrained to
known scattering amplitude as function of the energy. 
The method we propose enables an exact reproduction of the on-shell
\textit{NN} scattering data within an energy domain up to 
an upper limit.
There are no limitations regarding the presence of inelasticities 
above meson production threshold.
The resulting form factors are model- and energy-independent.
In order to keep the focus on the framework, its implementation
and scope, we limit this article to spin-uncoupled \textit{NN} states.

Separable representations of the \textit{NN} interaction have  the
advantage over their fundamental counterparts,
in that the number of operations needed to evaluate 
certain matrix elements in few- and many-body systems
diminishes drastically.
This sole element may dictate whether a computationally
intensive calculation becomes feasible or not under current 
floating-point operations per second 
({\footnotesize FLOPS}) speeds.
However, one has to keep in mind that separable interactions are 
artifacts designed to model specific aspects of the interaction.

Among the earliest developments, within inverse scattering theory,
is the possibility of constructing nonlocal separable interactions
extracting their form factors from phase-shift 
data~\cite{Yamagushi1954a,Bolsterli1965,Tabakin1969}. 
The particularity of this approach is that a rank-one separable
interaction can be obtained through inverse scattering with the 
only constrain given by the elastic scattering amplitude.
The method can be extended to cases where the phase shift crosses
the zero axis within the range of construction.
In Ref.~\cite{Kwong1997} this inverse-scattering approach was 
implemented for nucleon beam energies of up to 1.6~GeV, 
neglecting inelasticities~\cite{Kwong1997}.
Along a different line, in Ref.~\cite{Haidenbauer1984} a rank-$N$
separable potential was developed for the Paris 
potential~\cite{Lacombe1980}. 
The construction considers up to four Yukawa-type form factors,
providing optimal fits to \textit{NN} phase-shift data up to
300~MeV.
In Ref.~\cite{Bondarenko2011}, the authors have constructed a complex
separable potential to describe the 
$^3\textrm{SD}_1$ state fitted to data up to 1.1~GeV.
More recently, in Ref.~\cite{Khokhlov2021} the authors present
an algebraic method to solve inverse scattering based on Marchenko
theory, obtaining an energy-independent separable potential
in the $^1\textrm{S}_0$ channel up to 3~GeV. 

\begin{comment}
\end{comment}

This article is organized as follows.
In Sec. 2 we outline the theoretical framework where we construct
an inverse-scattering multi-rank 
separable potential from on-shell scattering amplitude.
We also discuss features exhibited by
the \textit{NN} phase-shift data relevant for this work.
This is followed with actual applications, where we present
solutions to channels of interest, paying attention on the 
structure of the solutions and the role of absorption in the data.
In Sec. 3 we summarize the main findings of this work
and discuss its scope.
\section{Theoretical framework}

\begin{comment}
As stated before, we aim to the construction of separable potentials
to be added to realistic ones in order to reproduce exactly
complex phase-shift data above pion production threshold.
For clarity in the presentation we find convenient to 
pursue the construction of separable potentials  addressing
three scenarios with increasing complexity:
(a)
rank-one pure separable potential for channels with non 
vanishing on-shell $t$-matrix;
(b) rank-$n$ pure separable potential for channels where
the on-shell $t$ matrix vanishes; and
(c) superposition of reference (realistic) potential with
rank-$n$ separable term.

If we assume colliding nucleons of equal mass $m$, then
the kinetic energy in the center-of-momentum (c.m.) reference frame 
at the pion threshold satisfies $2\sqrt{k^2\!+\!m^2}\!-\!2m\!=m_\pi$, 
with $k$ the relative momentum in the c.m. reference frame
and $m_\pi$ the pion mass.
This condition yields $k\!\approx\!1.8$~fm$^{-1}$, 
resulting in nucleon beam energy $E_{_L}\!\approx\!290$~MeV.
Typical upper energies of validity of high precision realistic
\text{NN} potential models range between 290 and 320~MeV.
\end{comment}

Let us consider two nucleons interacting by means of potential $\hat V$.
The equation of motion for the interacting pair in the
center-of-momentum reference frame (c.m.) is taken 
as~\cite{Kamada1998,Kamada2007}
\begin{equation}
  \label{schro}
  \left ( {\hat p^2} + m\hat V \right ) \Psi = {k^2}\Psi\;,
\end{equation}
where $m$ is the nucleon mass and
$k$ is the relativistic c.m. relative momentum for asymptotic states.
This approach has been broadly adopted by various groups
in the construction of high precision realistic \textit{NN} 
potentials~\cite{Wiringa1995,Lacombe1980,Stocks1994,Machleidt2001}.

The scattering $\hat T$ matrix associated to Eq.~\eqref{schro} reads 
\begin{equation}
\label{tv}
\hat T(z) = \hat V + \hat V\hat G_0(z) \hat T(z) \, ,
\end{equation}
with $\hat G_0(z)=m/(z-\hat p^2)$, the free propagator and
$z$ an energy parameter. 
For outgoing scattering states we express $z\!=\!\omega\!+\!i\eta$, 
with $\omega\!>\!0$, and $\eta$ a positive infinitesimal.
With this notation we denote
$G_0^{(+)}(\omega)\!=\!G_0(\omega+i\eta)$.

Considering a rank-one separable potential given by
\begin{equation}
  \label{defV}
  \hat V=|a\rangle \lambda \langle\tilde a|\;,
\end{equation}
with $\lambda$ a sign constant,
its corresponding scattering $T$ matrix becomes
\begin{equation}
  \label{rank1}
  \hat T(\omega) = 
  \frac{|a\rangle\lambda\langle\tilde a|}
       {1-\lambda\langle\tilde a|G_0^{(+)}(\omega)|a\rangle}
  \;.
\end{equation}
This equation establishes a link between the scattering matrix
$\hat T$, 
and the form factors $| a \rangle$ and $\langle\tilde a |$.
To proceed further we introduce the following ansatz for the 
form factors projected in momentum space
\begin{equation}
\label{ffactor}
\langle p| a\rangle =
  \langle\tilde a |p\rangle = \frac{f(p)}{\sqrt{mp}}\;,
\end{equation}
where $f(p)$ is a function to be determined from the data.
As we shall see, 
in the presence of absorption this function becomes complex.
Projecting Eq.~\eqref{rank1} in momentum space
with $\langle k|$ on the left, $|k\rangle$ on the right and
taking $\omega\!=\!k^2$, we obtain
\begin{equation}
\label{unfold}
t(k)\, 
\left [
1 - \frac{2}{\pi} \int_0^\infty 
\frac{\lambda f^2(p)\, p\,dp}{k^2+i\epsilon-p^2} 
\right ]
 = \lambda\,f^2(k)\;.
%= \frac{\lambda\,f^2(k)}{mk}\;,
\end{equation}
Here $t(k)\equiv mk\langle k|\hat T(k^2)|k\rangle$,
denotes the on-shell scattering matrix element. 
What turns out interesting from this result is that
for a given set of data defined by $t(k)$, 
there is a unique solution for $\varphi(k)\equiv\lambda f^2(k)$.
This feature is evidenced after discretizing the 
above integral in the momentum variable $p$, 
resulting in a linear equation for $\varphi(p)$.

To make contact with the scattering data
we adopt the parametrization of the scattering 
amplitude given in Refs.~\cite{Arndt1982,Arndt1983}, 
which for uncoupled channels is summarized by the $\bar K$ matrix
expressed as
\begin{equation}
\label{kmatrix}
  \bar K(k) = \tan\delta(k) + i\,\tan^2[\rho(k)]\;.
\end{equation}
Here $\delta$ and $\rho$ denote phase shift and absorption, 
respectively.
This matrix is related to $t$ through
\begin{equation}
\label{ttilde}
  t(k) = -\frac{\bar K(k)}{1 - i\bar K(k)}\;.
\end{equation}
With these definitions Eq.~\eqref{unfold} reduces to the following
linear equation for $\varphi$
\begin{equation}
\label{master}
  \bar K(k) \left [
1 - \frac{2}{\pi} {\cal P}\int_0^\infty 
\frac{\varphi(p) p\,dp}{k^2-p^2} 
  \right ]
 = -\varphi(k)\;,
\end{equation}
where ${\cal P}$ denotes principal-value integral.
This equation summarizes the interrelation between the square of
the separable form factor, $\varphi(k)$, and the on-shell 
scattering data contained in $\bar K$.
In the absence of inelasticities, when $\rho\!=\!0$, 
$\varphi$ becomes real. 
From the above equation we note that there is a one-to-one
correspondence between zeros of $K$ and those of
$\varphi$, meaning that
$\varphi$ keeps the same sign on intervals where $K$ does not
vanish.
Thus, to extract the form factor $f\!=\!\sqrt{\lambda\varphi}$,
we define $\lambda\!=\!\textrm{Sgn}\{\varphi\}$.

A necessary condition for the existence of a 
solution to Eq.~\eqref{master} is that for increasing $k$,
$\varphi(k)$ decreases at least as $\sim\!k^{-\nu}$, with $\nu\!>\!0$. 
Since the scattering data is given over a finite energy range,
there is no feasible way to verify this condition. 
A way to circumvent this limitation,
without compromising the ability to reproduce the data within 
its domain of definition, 
is by introducing a smooth cutoff to suppress high momentum 
contribution in the ${\cal P}$ integral.
Techniques of similar nature are adopted in the context of
renormalization group models~\cite{Furnstahl2012}.

In order to unfold $\varphi$ from Eq.~\eqref{master} we apply a 
smooth ultraviolet (UV) cutoff to  $\bar K(k)$, namely 
\begin{equation}
  \label{cutoff}
\bar K(k)\to \hat\Theta(k_c-k)\bar K(k)\,, 
\end{equation}
where
\begin{equation}
  \label{regulator}
  \hat\Theta(x) = \left \{
  \begin{array}{rl}
    1\,, &\hspace{10pt} \textrm{if $x\le 0$\;;} \\
    e^{-x^2/d^2}\,, &\hspace{10pt} \textrm{if $x>0$}\;;
  \end{array}
  \right .
\end{equation}
with $d$ its diffuseness.
Note that this regulator and its derivatives are continuous 
in the whole range. 
Its advantage is that it does not alter the integrand below $k_c$,
but it cuts off smoothly the high momentum components of $\bar K$.
The reason we apply the cutoff to $\bar K$ 
and not to the scattering amplitude is to preserve 
unitarity of the solutions when $\rho\!=\!0$. 
In this work we use $d\!=\!0.05$~fm$^{-1}$, throughout. 
\begin{comment}
Whenever $k_c\gtrsim 2$~fm$^{-1}$,
we affect mostly the very short range of the interaction, 
without altering the ability of the solution to reproduce 
the phase-shift data at the lower energies. 
\end{comment}

\subsection{Phase-shift data}
We base the present study on the SAID-SP07 
solution of the phase-shift analysis~\cite{Arndt2007} 
available at the INS Data Analysis Center of The George Washington 
University~\cite{said}, 
for $np$ data up to 2.5~GeV lab energy.
In the case of \textit{NN} uncoupled states
the data is represented by means of the phase shifts $\delta$ and
absorption $\rho$ over a broad range of laboratory energies $E$, 
with $k=\sqrt{mE/2}$, the relativistic relative momentum. 
Given that in our approach the nature of inverse-scattering
solutions depend on features of the scattering amplitude as 
function of the energy, 
we briefly examine them in the case of \textit{NN} uncoupled channels.

%namely $k\lesssim 1.8$~fm$^{-1}$.

In Table~\ref{tab:nnstates} we list all \textit{NN} uncoupled
states considered in this work.
Underlined states denote states with phase-shift crossing the zero axis 
at some energy below 2.5~GeV, 
property that requires specific considerations when there is no absorption.
States with nonzero $\rho$ are shown with gray background. 
As observed, the only elastic state  with nonzero phase-shifts are
$^1\textrm{F}_3$,
$^3\textrm{G}_4$,
$^1\textrm{H}_5$,
$^3\textrm{I}_6$ and
$^1\textrm{J}_7$.

\begin{table}[h]
%\begin{center}
    \begin{tabular}{|c|cccc cccc}
         \hline
      \rule[-6pt]{0pt}{18pt}
      $ST$ & \multicolumn{8}{c}{Allowed \textit{NN} states}\\
         \hline
      \rule[-4pt]{0pt}{18pt}
      0   1 & \cellcolor{gray!30}{\uline{\!$^1$S$_0$\!}} & $\cdot$ & 
              \cellcolor{gray!30}{\uline{\!$^1$D$_2$\!}} & $\cdot$ &
              \cellcolor{gray!30}{{\!$^1$G$_4$\!}} & $\cdot$ & 
              \cellcolor{gray!30}{{\!$^1$I$_6$\!}} & $\cdot$        \\
      \rule[-4pt]{0pt}{14pt}
      0   0 & $\cdot$ & 
               \cellcolor{gray!30}{\uline{{\!$^1$P$_1$\!}}} & $\cdot$ & 
                {\cellcolor{gray!0}{\!$^1$F$_3$\!}} & $\cdot$ & 
                {\cellcolor{gray!0}{\!$^1$H$_5$\!}} & $\cdot$ & 
                {\cellcolor{gray!0}{\!$^1$J$_7$\!}}             \\
      \rule[-4pt]{0pt}{14pt}
      1   0 & $\cdot$ & $\cdot$ & 
              \cellcolor{gray!0}{\uline{\!$^3$D$_2$\!}} & $\cdot$ &
              \cellcolor{gray!0}{\!$^3$G$_4$\!} &  $\cdot$  & 
              \cellcolor{gray!0}{\!$^3$I$_6$\!} &  $\cdot$  \\
      \rule[-8pt]{0pt}{14pt}
      1   1 &  \cellcolor{gray!30}{\uline{\!$^3$P$_0$\!}} & 
               \cellcolor{gray!30}{{\!$^3$P$_1$\!}} & $\cdot$ & 
               \cellcolor{gray!30}{\uline{\!$^3$F$_3$\!}} & $\cdot$ & 
               \cellcolor{gray!30}{{\!$^3$H$_5$\!}} & $\cdot$ & 
               \cellcolor{gray!30}{{\!$^3$J$_7$\!}} \\
         \hline
\end{tabular}
\caption{\label{tab:nnstates}
  \textit{NN} uncoupled states considered in this work.
  States with zeros in the phase-shift appear underlined.
  Those with absorption above pion threshold
  are shown on gray background.
  Features based on SAID SP07 phase-shift analysis of Ref.~\cite{said}.
  }
%\end{center}
\end{table}

\begin{figure}[ht]
  \begin{center}
\includegraphics[width=0.95\linewidth]{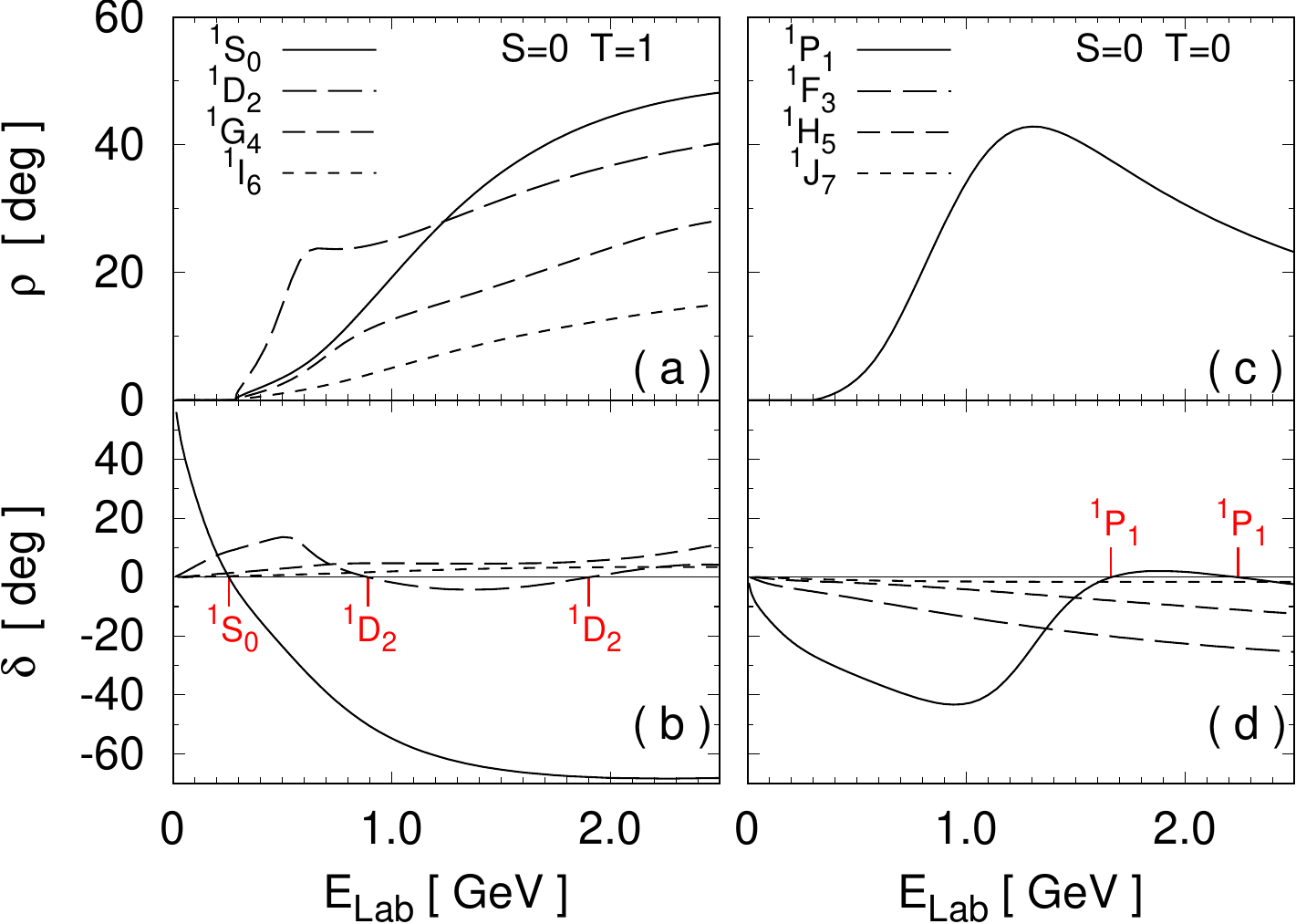}
\caption{
\label{fig:spin0}
  Absorption parameter $\rho$ [(a), (c)] 
  and phase-shift $\delta$ [(b), (d)]
  as functions of the nucleon beam energy $E_{Lab}$ for
  spin-0 \textit{NN} states with $J\!\leq\!7$. 
  Frames (a) and (b) correspond to $T\!=\!1$ channels, whereas
  frames (c) and (d) correspond to $T\!=\!0$.
  Vertical red segments indicate zeros in the phase-shift.
}
  \end{center}
\end{figure}

In Fig.~\ref{fig:spin0} we plot the phase-shift and absorption for
spin-zero states as functions of nucleon beam energy.
Panels (a) and (c) show the absorption parameter $\rho$ 
as a function of the nucleon beam energy $E_{Lab}$,
while panels (b) and (d) show the phase-shift $\delta$.
Frames (a) and (b) correspond to isovector ($T\!=\!1$) channels, whereas
frames (c) and (d) correspond to isoscalar ($T\!=\!0$) ones.
Short red segments indicate energies at which $\delta$ crosses
the zero axis in the $^1\textrm{S}_0$ and $^1\textrm{D}_2$ channels.
As observed, all isovector channels exhibit absorption above
the pion production threshold ($E_{Lab}\!\gtrsim\!290$~MeV).
Below 1~GeV the absorption is more pronounced for
the $^1\textrm{D}_2$ state.
We also notice that the phase-shifts $\delta$ vanishes
near 290~MeV in the case of $^1\textrm{S}_0$ state, and two times 
(around 890 and 1900~MeV) in the case of channel $^1\textrm{D}_2$. 
For the isoscalar channel, only $^1\textrm{P}_1$ state exhibits 
absorption, with zero phase-shift around 1660 and 2240~MeV.

\begin{figure}[ht]
\begin{center}
\includegraphics[width=0.95\linewidth]{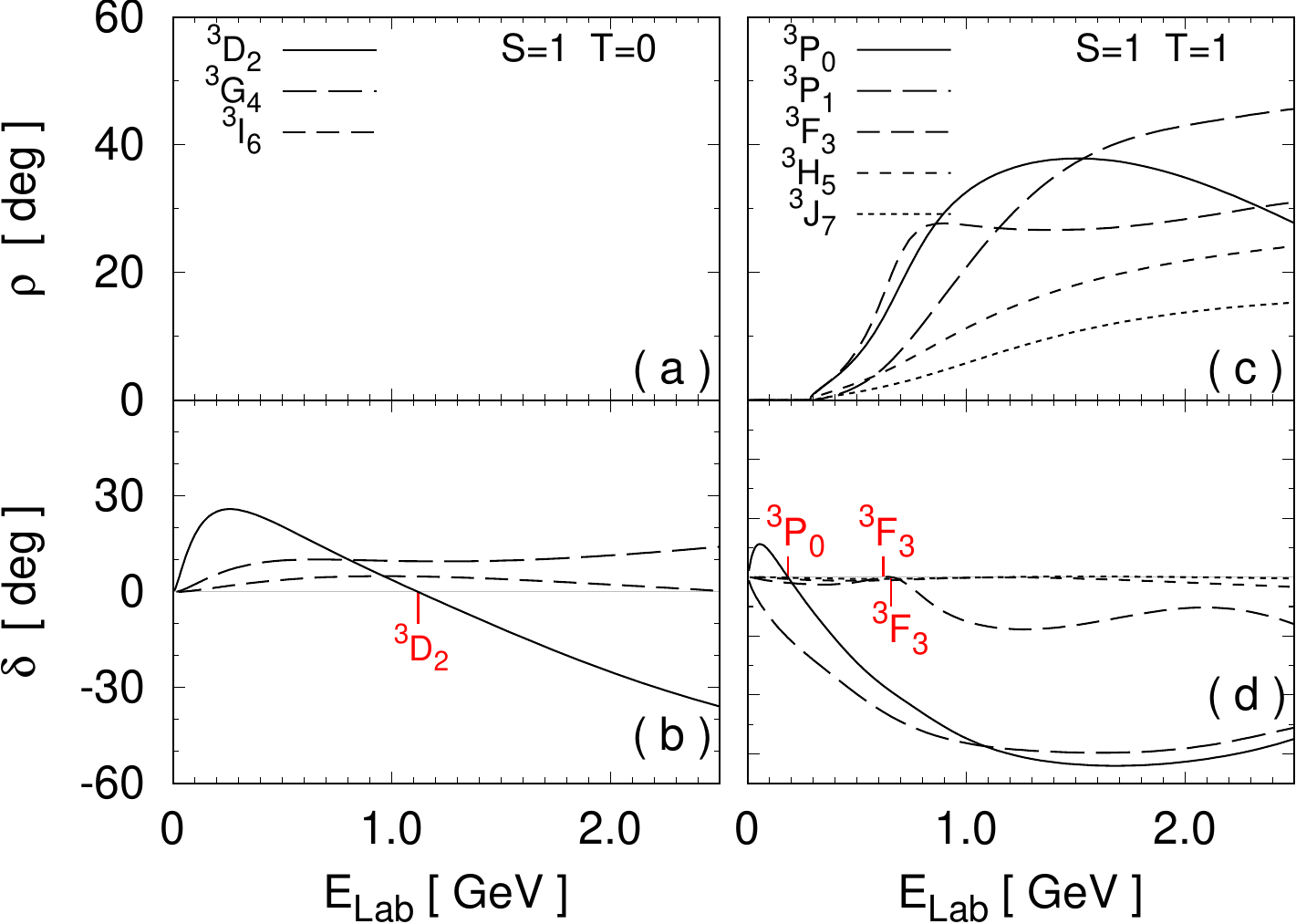}
\caption{
\label{fig:spin1}
  The same as Fig.~\ref{fig:spin0} but for spin-1 uncoupled
  \textit{NN} states.
}
  \end{center}
\end{figure}

In Fig.~\ref{fig:spin1} we show plots for $\delta$ and $\rho$
for $S\!=\!1$, spin-uncoupled states.
Frames (a) and (b) correspond to isoscalar channels, whereas
frames (c) and (d) correspond to isovector ones.
In the case of isoscalar states we observe that 
$^3\textrm{D}_2$, 
$^3\textrm{G}_4$ and
$^3\textrm{I}_I$ channels are elastic, 
while only $^3$D$_2$ exhibits zero phase-shift
around \num{1115}~MeV.
This scenario contrasts with the isovector channels,
where absorption is present in all states.
Below 1~GeV, the absorption is stronger
in the $^3\textrm{P}_0$ and $^3\textrm{P}_1$ states.
Additionally, the phase-shift in channel $^3\textrm{P}_0$ vanishes
around 200~MeV while for channel $^3\textrm{F}_3$ it crosses
twice the zero axis near 640~MeV.

\subsection{Solutions}
Based on the SP07 phase-shift analysis discussed in the preceding section,
we proceed to calculate the form factors $f(k)$ solutions of
Eq.~\eqref{unfold}, making use of
on-shell scattering data summarized by the matrix $T(k)$.
To solve this equation we discretize the momentum $p$ 
over a uniform array of $N$ elements, $p\to p_j=jh$, 
with $h\!=\!p_{max}/N$, the momentum step length.
Here $p_{max}\!=\!k_c\!+\Delta$, with $\Delta\approx 2$~fm$^{-1}$.
The resulting equation for $\varphi(p_j)=\varphi_j$ 
takes the matrix form
\begin{equation}
  \label{discrete}
   \sum_{j=1}^{N} \mathbbm{M}_{ij} \varphi_j = \mathbbm{b}_i\;,
\end{equation}
where $\mathbbm{M}$ is a square matrix built from on-shell data
and trapezoidal quadrature coefficients. 
The same holds for array $\mathbbm{b}$.
Regarding the low-momentum extrapolation of phase-shift data
we consider two cases:
\textit{a)}
$S\!=\!0$ waves, 
where we use the low-$k$ effective-range expansion; and
\textit{b)}
$S\!\geq\! 1$ waves, 
where we apply a power fit of the type $\delta(k)\!=\!a k^p$.
Once $\varphi$ is obtained by direct matrix inversion, 
we proceed to take its square root 
to obtain the form factor, $f(k)\!=\!\sqrt{\lambda\varphi(k)}$, 
with $\lambda$ the sign of the solution.

\begin{figure}[ht]
  \begin{center}
\includegraphics[width=0.9\linewidth]{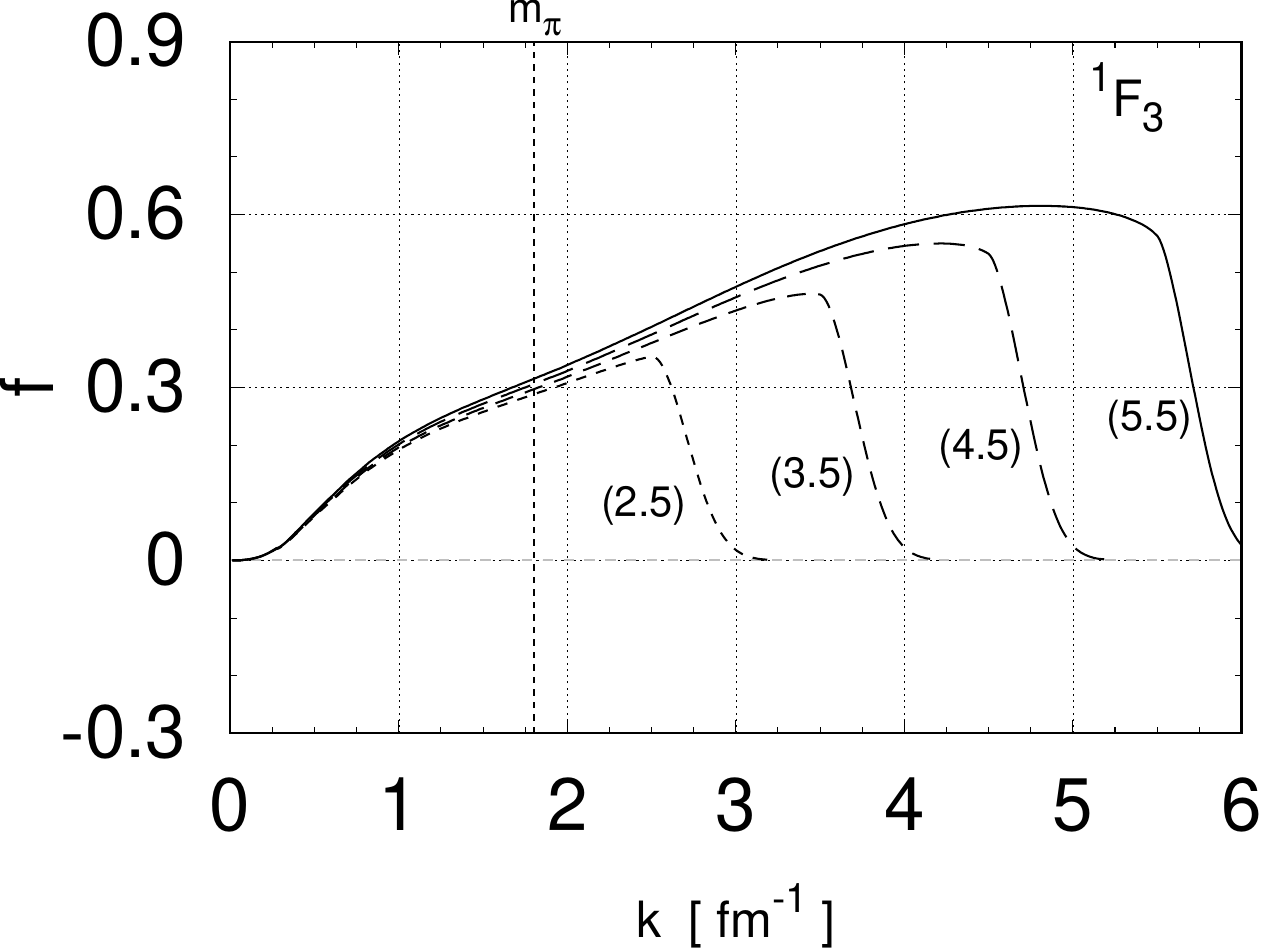}
\caption{
\label{fig:F1f3}
  Calculated form factor $f(k)$ as function of the relative
    momentum in the case of $^1\textrm{F}_3$ channel.
  Solid, long-, medium- and short-dashed curves denote cutoffs of
  \num{5.5}, \num{4.5}, \num{3.5} and \num{2.5}~fm$^{-1}$, respectively.
}
  \end{center}
\end{figure}

\begin{figure}[ht]
  \begin{center}
\includegraphics[width=0.9\linewidth]{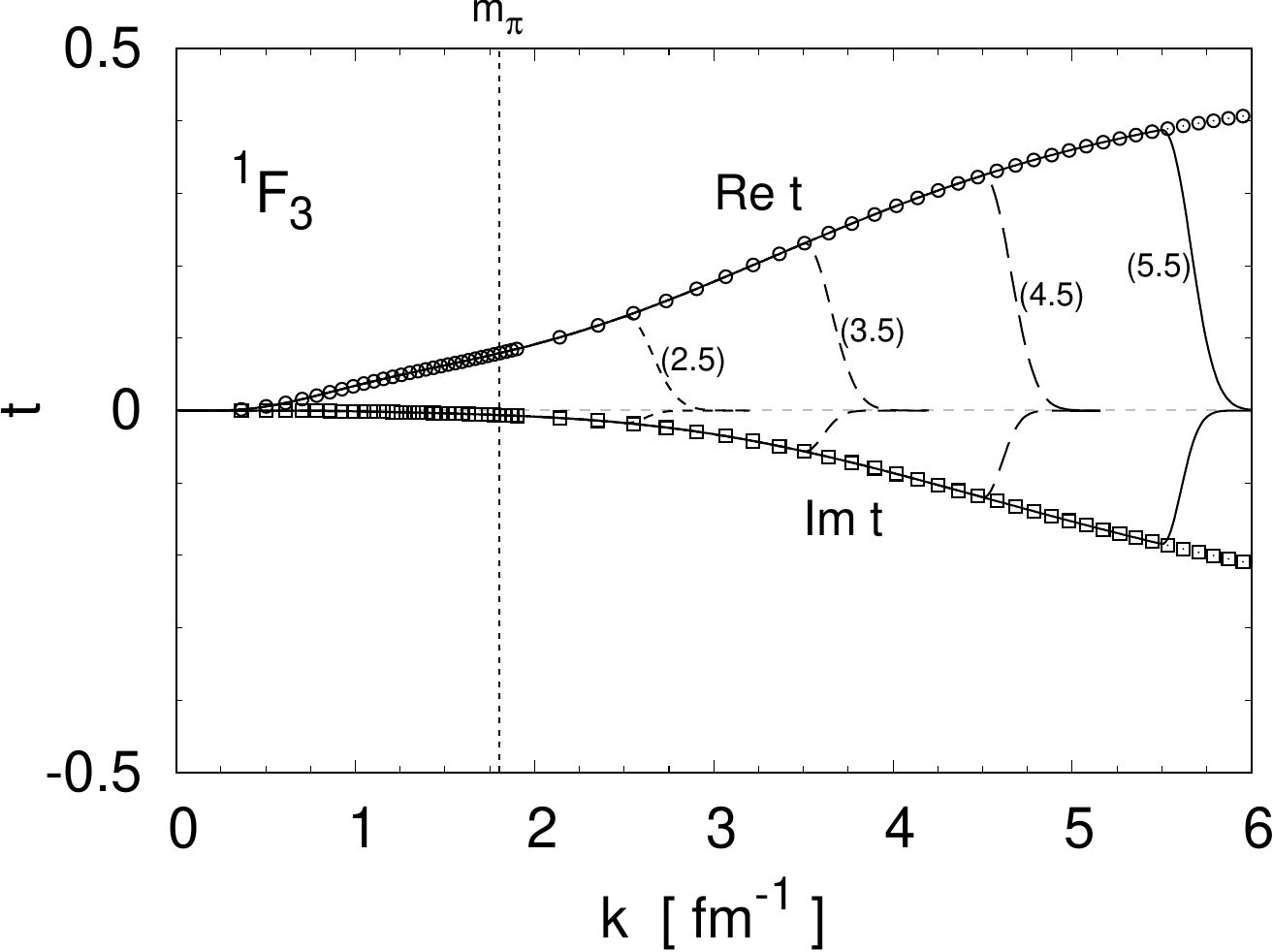}
\caption{
\label{fig:T1f3}
  $t$ matrix as function of the relative
    momentum for the $^1\textrm{F}_3$ channel.
    Circles (squares) denote the real (imaginary) component of $t$
    based on the SP07 data.
  Solid, long-, medium- and short-dashed curves denote $t$-matrix based
    on separable solutions with UV cutoffs at
  \num{5.5}, \num{4.5}, \num{3.5} and \num{2.5}~fm$^{-1}$, respectively.
}
  \end{center}
\end{figure}

The procedure outlined above is applied to the
$^{1}\textrm{F}_{3}$ channel, featuring no absorption 
and nonzero phase shift in the whole energy range.
To illustrate the role of the UV cutoff, we consider
four different values:
$k_c\!=\!2.5$, 3.5, 4.5 and 5.5~fm$^{-1}$.
The corresponding results for $f(k)$ are shown in Fig.~\ref{fig:F1f3}, 
where labels in parenthesis indicate $k_c$ in fm$^{-1}$ units.
These energy-independent solutions define rank-one separable potentials
reproducing exactly the phase-shift $\delta(k)$ up to $k\!=\!k_c$.
This is illustrated in Fig.~\ref{fig:T1f3},
where we plot the on-shell $t$ matrix as given by the phase-shift
as function of $k$, the relativistic relative momentum in the c.m.
Open circles and squares denote Re$\{t\}$ and Im$\{t\}$,
respectively.
%==================================================================
%==================================================================
The calculated $t$ matrix for the separable potential
is obtained from Eq.~\eqref{rank1}.
As observed, the agreement between the data and the $t$ matrix
from the inverse-scattering form factors is quite good up to
$k\!=\!k_c$. 
Above this bound Re$\{t\}$ and Im$\{t\}$ fall rapidly due
to the regulator.
%==================================================================
%==================================================================

\subsubsection{Solutions for $^1\textrm{S}_0$ channel.-}
\label{sec:f1s0}
Let us now address the $^1\textrm{S}_0$ channel, 
featuring a zero in $\delta(k)$ in the energy range. 
Channels with similar behavior to this one are
$^1\textrm{D}_2$, $^3\textrm{D}_2$, $^3\textrm{P}_0$ and 
$^3\textrm{F}_3$.
In principle we can also apply Eq.~\eqref{discrete} to obtain
the corresponding $\varphi(k)$. The resulting solution
also exhibits a zero at the point at which $\delta$ vanishes.
In the particular case when $\rho\!=\!0$, 
this leads to a real and continuous $\varphi(k)$, with a node
at $k\!=\!k_1$, the zero of the phase-shift.
Therefore, the resulting $f(k)$ becomes complex, an scenario we choose
to avoid.

The effect of a node in the phase-shift has been addressed
by Bolsteri in Ref.~\cite{Bolsterli1965}, proposing
the construction of piecewise solutions over
intervals where the phase-shift does not have a zero.
Following this idea and considering that the phase-shift
in channel $^1\textrm{S}_0$ has a single
zero at a given momentum $k_1$, we introduce the rank-two 
interaction
\begin{equation}
  \label{rank2}
  \hat V=|a_1\rangle\lambda_1\langle\tilde a_1| +
         |a_2\rangle\lambda_2\langle\tilde a_2|\;,
\end{equation}
with $\lambda_2\!=\!-\lambda_1$.
We also define
\begin{subequations}
\begin{eqnarray}
\label{ffactor_a1}
\langle k|a_1\rangle &=\langle\tilde a_1|k\rangle=
  \Theta(k_1-k)\, \displaystyle{\frac{f_1(k)}{\sqrt{mk}}}\;;
   \\
\label{ffactor_a2}
\langle k|a_2\rangle &=\langle\tilde a_2|k\rangle=
  \Theta(k-k_1)\,\displaystyle{\frac{f_2(k)}{\sqrt{mk}}} \;;
\end{eqnarray}
\end{subequations}
where $\Theta$ denotes the Heaviside step function while
$f_1$ and $f_2$ are form factors associated with intervals
$[0,k_1]$ and $[k_1,k_{max}]$, respectively.
With these definitions we note that
$\langle \tilde a_1|a_2\rangle\!=\!\langle\tilde a_2|a_1\rangle\!=\!0$.

To obtain the equations for $\varphi_1=f_1^2$, and $\varphi_2=f_2^2$,
the procedure outlined in the Appendix yields
\begin{subequations}
\begin{eqnarray}
\label{unfold01}
  \bar K(k) \left [
  1 - \frac{2}{\pi} {\cal P}\int_0^{k_1} 
\frac{\varphi_1(p) p\,dp}{k^2-p^2} 
  \right ]
  =& -\varphi_1(k)\,; \\
\label{unfold02}
  \bar K(k) \left [
  1 - \frac{2}{\pi} {\cal P}\int_{k_1}^{\infty} 
\frac{\varphi_2(p) p\,dp}{k^2-p^2} 
  \right ]
  =& -\varphi_2(k)\,.
\end{eqnarray}
\end{subequations}
In Eqs.~\eqref{unfold01} and \eqref{unfold02} we impose
$k\!\leq k_1$, and
$k\!\geq k_1$, respectively.
Thus, the equations for inverse form factors $f_1$ and $f_2$ 
take the form Eq.~\eqref{discrete},
which are solved independently on each of the two intervals. 

\begin{figure}[ht]
  \begin{center}
\includegraphics[width=0.90\linewidth]{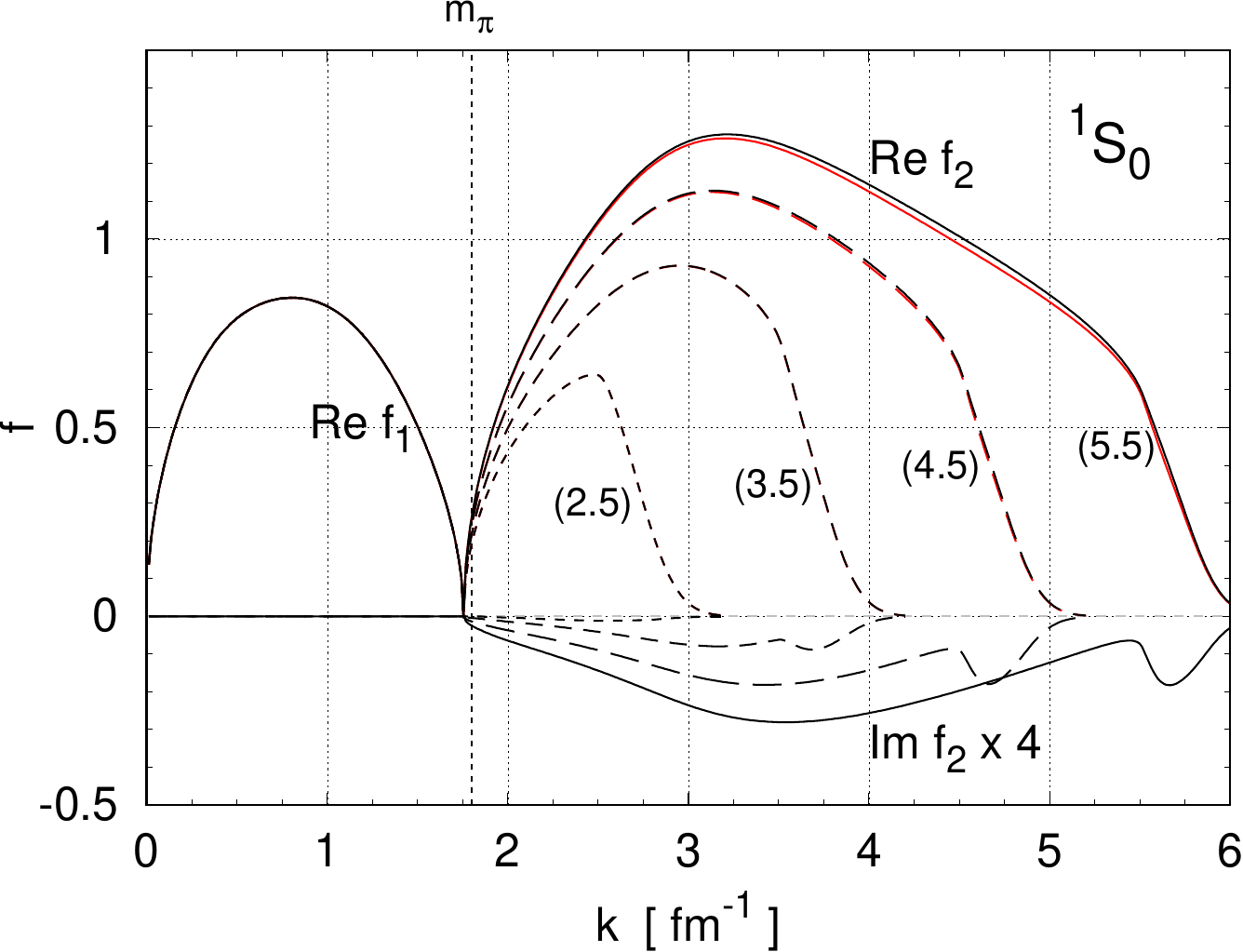}
\caption{
\label{fig:F1s0}
  Calculated form factors $f_1$ and $f_2$ as functions of the 
  relative momentum for the $^1\textrm{S}_0$ channel.
  Solid, long-, medium- and short-dashed curves denote cutoffs of
  \num{5.5}, \num{4.5}, \num{3.5} and \num{2.5}~fm$^{-1}$, respectively.
  Solutions with $\rho\!=\!0$, are shown in red, while those
  with $\rho\!\neq\!0$ are shown in black.
  The imaginary components of $f_2$ have been amplified by a factor of four.
}
  \end{center}
\end{figure}

In Fig.~\ref{fig:F1s0} we plot the calculated form factors
$f_1$ and $f_2$ as functions of the relative momentum for the
$^1\textrm{S}_0$ channel.
Red curves represent results suppressing absorption ($\rho\!=\!0$)
while black curves including it ($\rho\!\neq\!0$).
In both cases $\lambda_1\!=\!-1$, and $\lambda_2\!=\!+1$.
Solid, long-, medium- and short-dashed curves correspond
to results with $k_c\!=\!5.5$, \num{4.5}, \num{3.5} and 
\num{2.5}~fm$^{-1}$, respectively.
We observe mild differences in $\textrm{Re}\{f_2\}$ when absorption
is considered, being more pronounced when $k_c\!=\!5.5$~fm$^{-1}$.
Results for $\textrm{Im}\{f\}$, taking place for nonzero absorption, 
have been amplified by a factor of four ($\times 4$). 
As observed, these weak 
contributions have negative sign relative to their real counterpart.
%==================================================================
%==================================================================
%==================================================================

\begin{figure}[ht]
  \begin{center}
\includegraphics[width=0.9\linewidth]{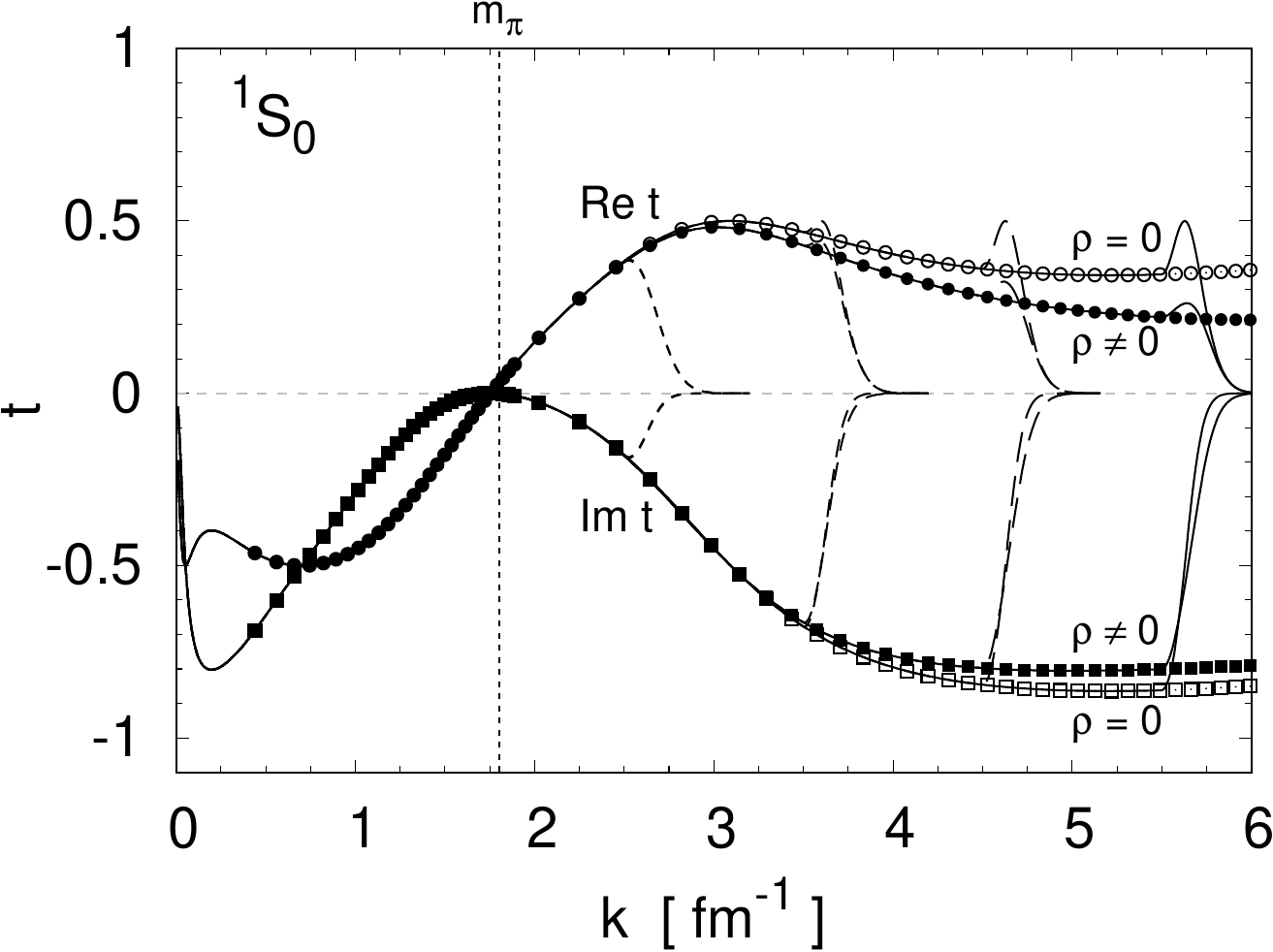}
\caption{
\label{fig:T1s0}
  $t$ matrix as function of the relative
    momentum for the $^1\textrm{S}_0$ channel.
    Circles (squares) denote the real (imaginary) component of $t$
    based on the SP07 data.
    Open (filled) symbols correspond to cases with $\rho\!=\!0$ 
    ($\rho\!\neq\! 0$).
    Continuous curves denote $t$ from the separable solutions
    with different UV cutoffs, adopting the same
    pattern convention as in Fig.~\ref{fig:F1s0}.
}
  \end{center}
\end{figure}

In Fig.~\ref{fig:T1s0} we compare the calculated $t$ matrix obtained
from inverse-scattering potentials and the data,
including the cases $\rho=0$, and $\rho\neq 0$.
Different curve textures are used for each UV cutoff, 
adopting the same convention as in Fig.~\ref{fig:F1s0}.
Circles (squares) denote $\textrm{Re}\{t\}$
($\textrm{Im}\{t\}$) from the data.
Filled and empty symbols denote inclusion and suppression of absorption,
respectively.
We observe that the inverse-scattering solutions match the data
in the whole $k$-range, up to their respective UV cutoff.
%==================================================================

\begin{figure}[ht]
  \begin{center}
\includegraphics[width=0.90\linewidth]{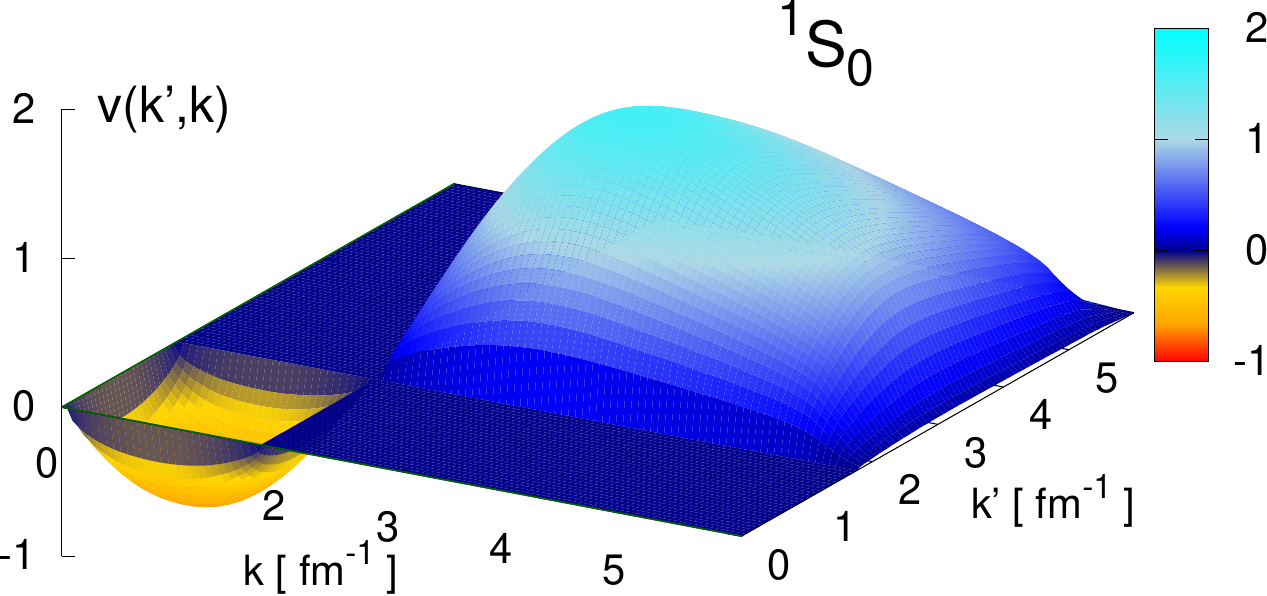}
\caption{
\label{fig:v1s0}
    Surface plot of the rank-2 energy-independent separable 
    potential $v(k',k)$ in the $^1\textrm{S}_0$ channel,
    as function of the relative momentum.
}
  \end{center}
\end{figure}

To visualize the shape of the inversion potential in the $kk'$ plane,
Fig.~\ref{fig:v1s0} shows a surface plot of 
$v(k',k)\!=\!\langle k'|\hat V|k\rangle$ for the rank-2 solution
in the $^1\textrm{S}_0$ channel. 
For this illustration we suppress absorption by setting $\rho\!=\!0$.
Here we observe a tile structure formed by two domes on square domains,
with nodal lines $k\!=\!k_1$, and $k'\!=\!k_1$,
taking place at the zero of the phase-shift.
The convex dome, taking place when $k$ and $k'$ lie below $k_1$, 
is determined by the positive phase-shift in this channel, 
as seen in panel (b) of Fig.~\ref{fig:spin0}.
By construction, this energy-independent potential accounts for 
all phase-shift up to $k\!=\!5.5$~fm$^{-1}$, 
namely 2.5~GeV lab energy.
%==================================================================

%==================================================================

\subsubsection{The case of $^1\textrm{P}_1$ channel.-}
\label{sec:f1p1}
The $^1\textrm{P}_1$ channel is the only singlet-odd channel
with absorption, having also two zeros in the phase-shift.
Following our previous discussion, if absorption is omitted, then
the separable potential is of rank 3, that is to say
\begin{equation}
  \label{rank3}
  \hat V=|a_1\rangle\lambda_1\langle\tilde a_1| +
         |a_2\rangle\lambda_2\langle\tilde a_2| +
         |a_3\rangle\lambda_3\langle\tilde a_3| \;,
\end{equation}
with $\lambda_1\!=\!\lambda_3\!=\!-\lambda_2$.
The corresponding form factors are in this case
\begin{subequations}
  \begin{align}
\label{ffactor_b1}
  \langle k|a_1\rangle =\langle\tilde a_1|k\rangle &=
  \Theta(k_1-k)\, \displaystyle{\frac{f_1(k)}{\sqrt{mk}}}\;;
   \\
\label{ffactor_b2}
  \langle k|a_2\rangle =\langle\tilde a_2|k\rangle &=
  \Theta(k-k_1)\,\Theta(k_2-k)\displaystyle{\frac{f_2(k)}{\sqrt{mk}}} \;;
   \\
\label{ffactor_b3}
  \langle k|a_3\rangle =\langle\tilde a_3|k\rangle &=
  \Theta(k-k_2)\,\displaystyle{\frac{f_3(k)}{\sqrt{mk}}} \;.
  \end{align}
\end{subequations}
Following the Appendix, the associated equations for 
$\varphi_1\!=\!f_1^2$,
$\varphi_2\!=\!f_2^2$ and
$\varphi_3\!=\!f_3^2$  become
\begin{subequations}
\begin{eqnarray}
\label{unfold1}
  \bar K(k) \left [ 1 - \frac{2}{\pi} {\cal P}\int_{0}^{k_1} 
\frac{\varphi_1(p) p\,dp}{k^2-p^2} \right ]
  =& -\varphi_1(k)\,; \\
\label{unfold2}
  \bar K(k) \left [ 1 - \frac{2}{\pi} {\cal P}\int_{k_1}^{k_2} 
\frac{\varphi_2(p) p\,dp}{k^2-p^2} \right ]
  =& -\varphi_2(k)\,; \\
\label{unfold3}
  \bar K(k) \left [ 1 - \frac{2}{\pi} {\cal P}\int_{k_2}^{\infty} 
\frac{\varphi_3(p) p\,dp}{k^2-p^2} \right ]
  =& -\varphi_3(k)\,.
\end{eqnarray}
\end{subequations}
To obtain the solutions $\varphi_1$, 
$\varphi_2$ and $\varphi_3$ we follow an analogous procedure to the
one applied for the $^1\textrm{S}_0$ channel.

\begin{figure}[ht]
  \begin{center}
\includegraphics[width=0.9\linewidth]{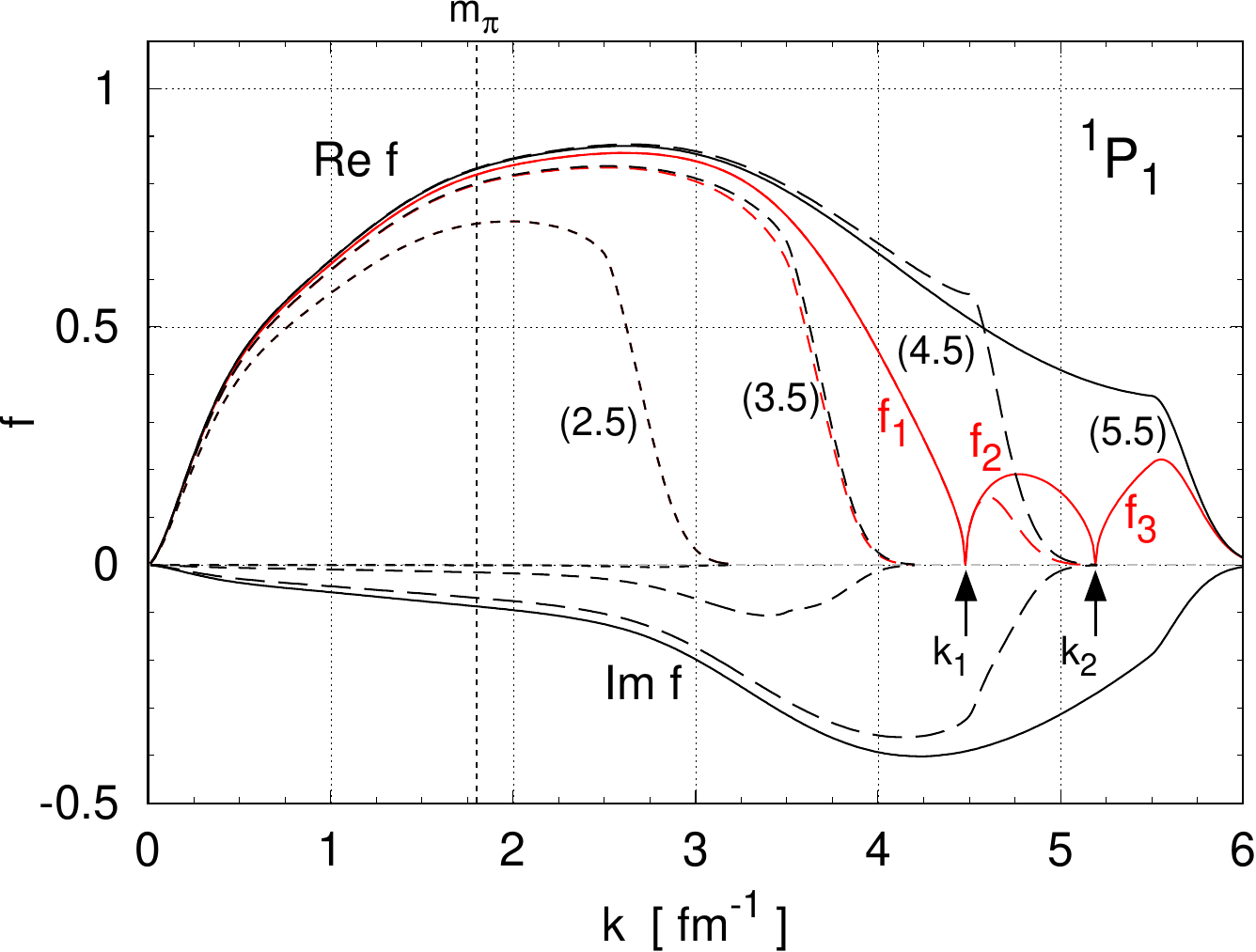}
\caption{
\label{fig:F1p1}
  Calculated form factor $f(k)$ as function of the relative
    momentum in the case of $^1\textrm{P}_1$ channel.
  Black and red curves represent results including and excluding
  absorption, respectively.
  Solid, long-, medium- and short-dashed curves denote cutoffs of
  \num{5.5}, \num{4.5}, \num{3.5} and \num{2.5}~fm$^{-1}$, respectively.
  Labels $k_1$ and $k_2$ correspond to zeros of the phase-shift.
}
  \end{center}
\end{figure}

In Fig.~\ref{fig:F1p1} we plot the resulting form factors $f(k)$ 
for channel $^1\textrm{P}_1$ as 
function of the relative momentum $k$.
Red curves denote solutions without absorption ($\rho\!=\!0$), 
displaying only real solutions.
Black curves denote solutions including absorption ($\rho\!\neq\!0$), 
with $\textrm{Re}\{f\}$ positive and $\textrm{Im}\{f\}$ negative.
Solid, long-dashed, dashed and short-dashed curves denote
solutions with UV cutoff at 5.5, 4.5, 3.5 and 2.5~fm$^{-1}$,
respectively.
The vertical blue line correspond to the pion mass.
%==================================================================
%==================================================================

We observe that all solutions $f(k)$ vanish at the origin
while they fall rapidly to zero above their respective $k_c$. 
Solid red curves denote the rank-3 solutions 
($f_1$, $f_2$ and $f_3$) without absorption, with $k_c\!=\!5.5$~fm$^{-1}$.
This rank arises from the two zeros of the phase-shift, $k_1$ and $k_2$, 
taking place below $k_c$.
As $k_c$ diminishes to 4.5~fm$^{-1}$, the rank of the solution
decreases to two (red long-dashed curves).
For $k_c\!=\!3.5$~fm$^{-1}$ and below, the solution becomes rank-1
(red dashed and short-dashed curves).
When absorption is accounted for, $\bar K(k)$ does not have zeros
in the whole range, leading to rank-1 solutions. 
These are shown with black curves whose patterns depend on $k_c$,
adopting the same convention as in Fig.~\ref{fig:F1s0}.
Focusing on the two solid curves, we notice that the solution with
absorption ($\rho\!\neq\!0$) wraps the rank-3 solutions in red.

\begin{figure}[ht]
\begin{center}
\includegraphics[width=0.9\linewidth]{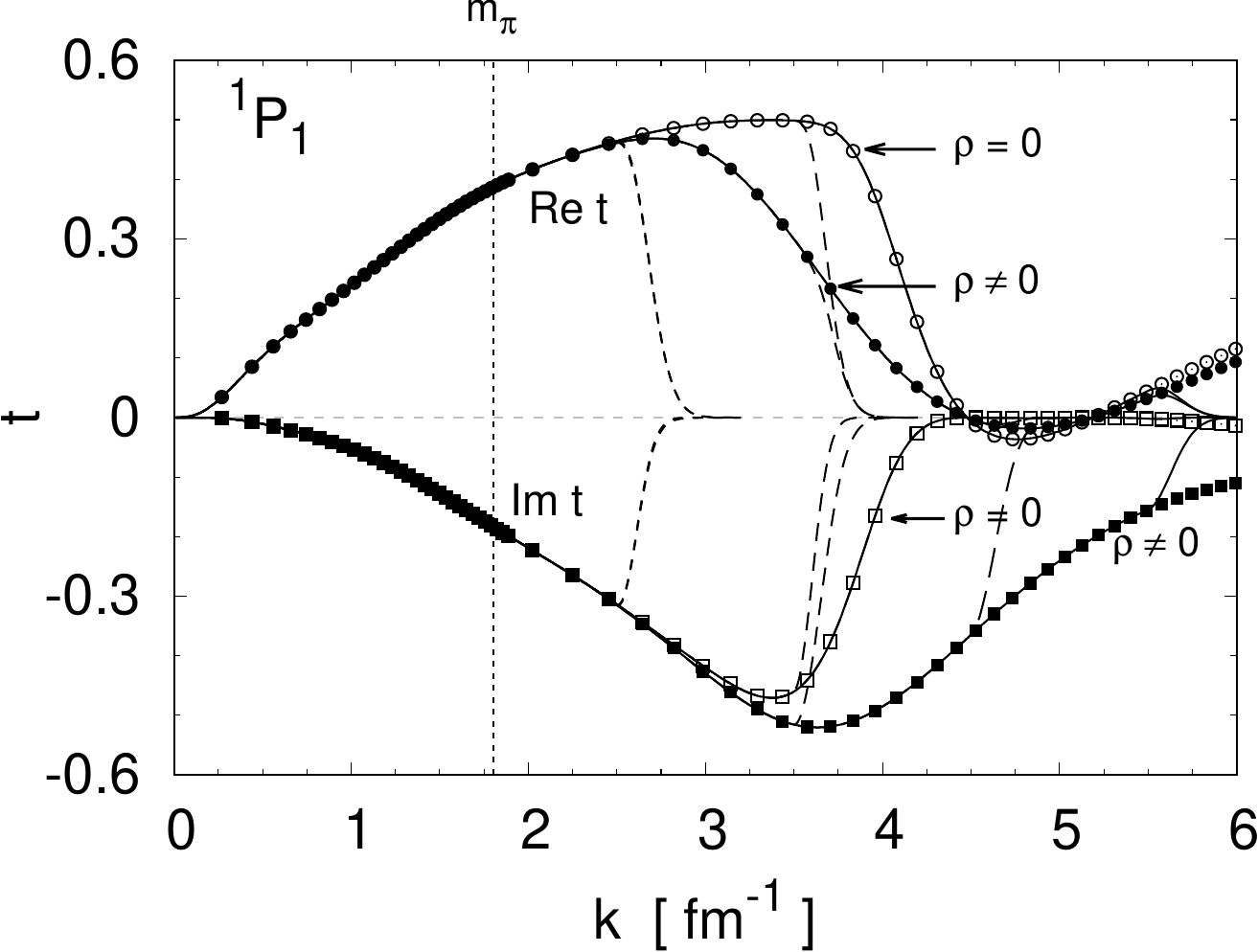}
\end{center}
\caption{
\label{fig:T1p1}
  On-shell $t$ matrix in the $^1\textrm{P}_1$ channel
  as function of the relative momentum.
  Circles (squares) denote $\textrm{Re}\{t\}$ 
  ($\textrm{Im}\{t\}$) from the data.
Filled and empty symbols denote inclusion and suppression of absorption,
respectively.
  Curve patterns follow the same convention as in Fig.~\ref{fig:F1s0}
}
\end{figure}

The ability of these inversion solutions to reproduce the on-shell
data is illustrated in Fig. \ref{fig:T1p1}, where we plot the
real and imaginary components of the on-shell $t$ matrix as functions
of the relative momentum $k$.
Circles (squares) denote $\textrm{Re}\{t\}$ 
($\textrm{Im}\{t\}$) from the data.
Filled and empty symbols denote inclusion and suppression of absorption,
respectively.
Different curve textures are used for each UV cutoff, 
adopting the same convention as in Fig.~\ref{fig:F1s0}.
We observe that the inverse-scattering solutions match the data
in the whole range of momenta, up to its respective UV cutoff $k_c$.
This level of agreement is replicated to all other states considered
in this study.

\subsubsection{Global results}
\begin{figure}[ht]
\includegraphics[width=0.95\linewidth]{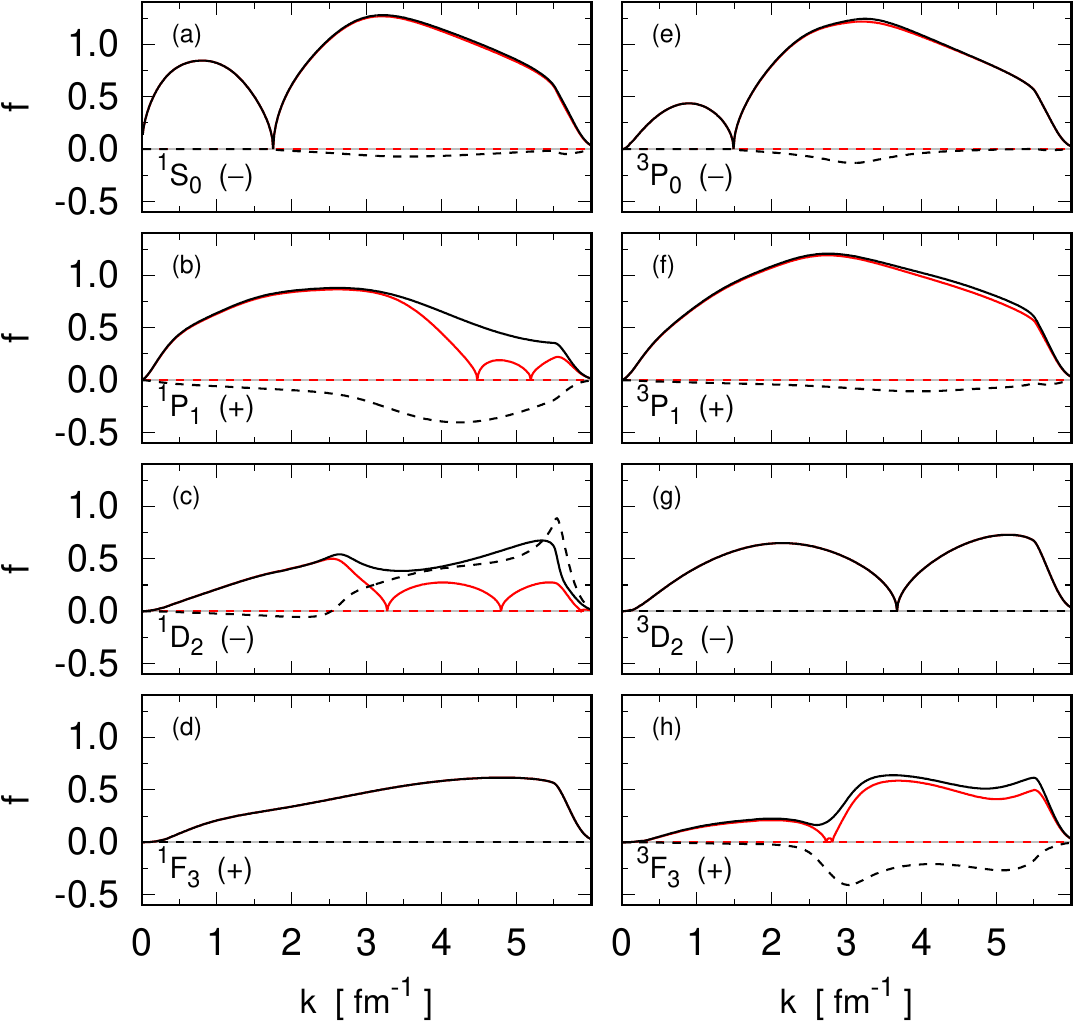}
\caption{
\label{fig:f0j3}
  Calculated form factor $f(k)$ as function of the relative
  momentum for \textit{NN} uncoupled states with $J\!\leq\!3$.
  Solid and dashed curves denote $\textrm{Re}\{f\}$
  and $\textrm{Im}\{f\}$,
  respectively.
  Black (red) curves denote solutions with $\rho\!\neq\!0$ 
  ($\rho\!=\!0$).
}
\end{figure}

We extend the applications described in the previous sub-sections
to all states with $J\!\leq\!7$.
The UV cutoff is kept equal to 5.5~fm$^{-1}$ throughout.
In Fig.~\ref{fig:f0j3} we plot the resulting form factors $f$
for all states with $0\!\leq\! J\!\leq\!3$.
Black and red curves denote results including and suppressing
absorption, respectively.
Solid and dashed curves represent
$\textrm{Re}\{f\}$ and $\textrm{Im}\{t\}$, respectively.
The $(\pm)$ signs on the lower left-and-side of each panel
denote the sign of $\lambda_1$ of the solutions.
Panels (a-d) show results for singlet states while
panels (e-h) show the ones for triplet states.
We note that the role of absorption in the solutions
appear most pronounced in
channels $^1\textrm{P}_1$, $^1\textrm{D}_2$ and $^3\textrm{F}_3$,
as differences between red and black curves become more evident.
%==================================================================
%==================================================================

\begin{figure}[ht]
\includegraphics[width=0.95\linewidth]{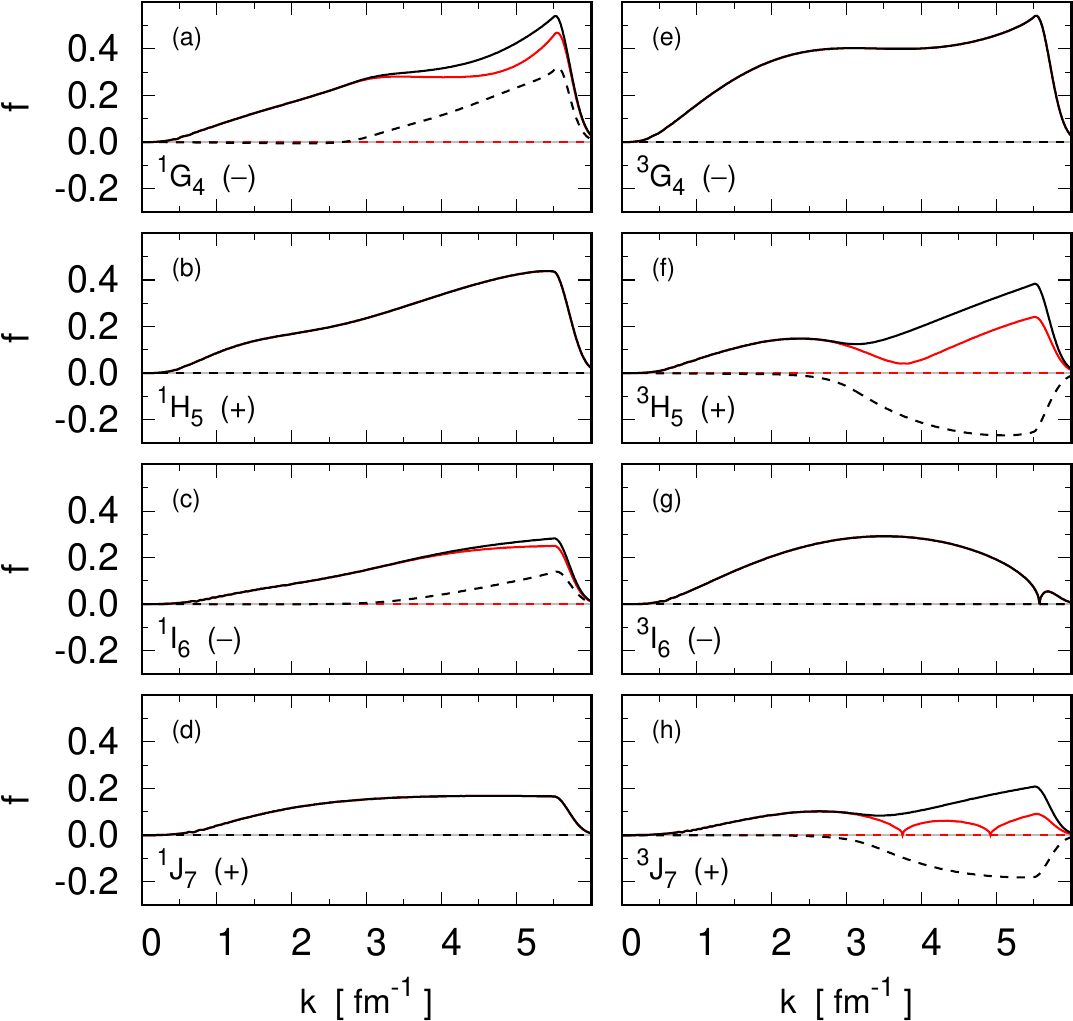}
\caption{
\label{fig:f4j7}
  Calculated form factor $f(k)$ as function of the relative
  momentum for \textit{NN} uncoupled states with $4\!\leq\!J\!\leq\!7$.
  Solid and dashed curves denote $\textrm{Re}\{f\}$ and 
  $\textrm{Im}\{f\}$, respectively.
  Black (red) curves denote solutions with $\rho\!\neq\!0$ 
  ($\rho\!=\!0$).
}
\end{figure}

In Fig.~\ref{fig:f4j7} we plot the resulting form factors $f$
for all states with $4\!\leq\! J\!\leq\!7$.
The curve colors and patterns follow the same convention as in Fig.~\ref{fig:f0j3}.
Panels (a-d) show results for singlet states while
panels (e-h) show the ones for triplet states.
In general all solutions follow the same trend as in the cases for $J\!\leq\!3$,
with non-negligible absorption in the triplet states $^3\textrm{H}_5$ and
$^3\textrm{J}_7$. 
%==================================================================
%==================================================================
%==================================================================

\begin{figure}[ht]
\includegraphics[width=0.95\linewidth]{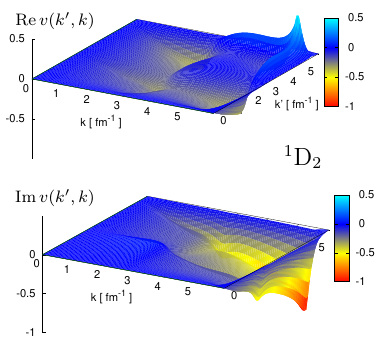}
\caption{
  \label{fig:3d_1d2}
    Surface plot of the rank-1 energy-independent separable 
    potential $v(k',k)$ in the $^1\textrm{D}_2$ channel,
    as function of the relative momentum.
  Upper and lower surfaces correspond to the real and
  imaginary components of the potential, respectively.
}
\end{figure}

To illustrate the non-Hermitian structure of the potential in the
$^1\textrm{D}_2$ state, in Fig.~\ref{fig:3d_1d2} we show a surface
plot of the resulting rank-1 separable potential in the $kk'$ plane.
Here $v(k',k)\!=\!\lambda f(k')f(k)$, with $\lambda\!=\!-1$.
The upper and lower surfaces represent the real and imaginary components
of the potential, respectively.
By observing $\textrm{Re}\{v\}$ we notice overall attraction
followed by strong repulsion at short distances (large $k$,$k'$).
Similarly, $\textrm{Im}\{v\}$ 
exhibits strong absorption at short distances.
The actual effect of these features in the \textit{NN} potential 
can be investigated in more specific applications, 
such as nucleon-nucleus scattering~\cite{Arellano2002},
an issue of interest but beyond the focus of this work.
%==================================================================

\section{Summary and conclusions}

  We have presented a method to construct, within inverse-scattering
  theory, an energy-independent separable potential capable
  of reproducing both phase-shift and absorption over a
  predefined energy range.
  The approach relies on the construction of non-overlapping
  multi-rank separable potentials, whose form factors satisfy 
  simple linear equations on intervals where the
  $K$ matrix does have zeros.
  The method is applied to \textit{NN} interactions with
  scattering data taken from the {\footnotesize SAID-SP07}
  phase-shift analysis with focus on spin-uncoupled states,
  considering a Schr\"odinger-like wave equation with
  minimal relativity.
  The inversion potentials are channel dependent,
  of varying rank depending on the number of zeros 
  of the $K$ matrix, reproducing exactly the data up to the
  selected upper momentum.

  The method we have introduced allows for the selective 
  inclusion/exclusion of absorption in the calculation of 
  form factors of the inversion potential.
  In this respect the framework is broad enough to assess
  implications of bare \textit{NN} interactions in domains 
  where loss of flux becomes important, 
  an issue of relevance in near-GeV nucleon-nucleus 
  collisions~\cite{Arellano2002}.

  Although applications made here have been
  restricted to Schr\"odinger-like wave equations, 
  other more general frameworks such as Thompson's relativistic 
  equation~\cite{Brockman1990,Thompson1970,Eyser2004} 
  can naturally be incorporated.
  In such a case the evaluation of
  $\langle\tilde a|\hat G_0^{(+)}(\omega)|a\rangle$
  would involve the energy denominator
  \begin{equation}
    \label{rel-prop}
    2\,(m^2+\omega)^{1/2}+i\epsilon-2\,(m^2+\hat p^2)^{1/2}\;,
  \end{equation}
  with $\omega\!=\!k^2$.
  The decay of the associated propagator as a function of 
  the intermediate momenta goes as $\sim\!1/p$.
  Thus, the use of UV cutoffs becomes instrumental 
  for obtaining an equation for $\varphi$, over a finite $p$-array,
  analogous to that in Eq.~\eqref{discrete}

  For clarity purposes we have focused this work on laying out
  the inversion formalism and investigate features of the solutions, 
  considering spin-uncoupled states
  as given by the \textit{NN} scattering data.
  Its extension to coupled states requires specific considerations,
  whose results will be reported elsewhere.
  The basic idea in this case 
  is to consider Eq.\eqref{tv} for coupled states,
  resulting in
  \begin{equation}
    \label {t2x2}
    \hat T(z) = \left [ 1 - \hat V \hat G_0(z)\right ]^{-1}\hat V\,.
  \end{equation}
  When projected on shell, $\langle k |\hat T(k^2)|k\rangle$
  becomes a $2\times 2$ matrix which can then be diagonalized 
  \textit{via} a $k$-dependent passage matrix. 
  The transformed $2\times 2$ potential can then be assumed
  diagonal with separable terms. In this way the problem
  gets reduced to the search of two separable solutions, 
  one for each term in the diagonalyzed $T$ matrix. 

\appendix
  \renewcommand{\theequation}{A-\arabic{equation}}
  % redefine the command that creates the equation no.
  \setcounter{equation}{0}  % reset counter
\section*{Appendix}
\subsection*{$T$ matrix for non-overlapping multi-rank separable
potentials}
\label{Appendix1}

Let us consider a non-overlapping rank-$N$ potential defined as
\begin{equation}
  \label{rank-n}
  \hat V = \sum_{i=1}^{N} |a_i\rangle \lambda_i \langle \tilde a_i|\;,
\end{equation}
where we impose
\begin{equation}
  \label{aij}
  \langle \tilde a_i|a_j\rangle = 0\;, \qquad\textrm{ for $i\!\neq\! j$.}
\end{equation}
In the context of this study, a way to achieve this non-overlapping feature
is by defining 
\begin{subequations}
\begin{align}
  \label{overlap1}
  \langle \tilde a_i|k\rangle &= 
        \Theta(k-k_{i-1})\Theta(k_{i}-k) \tilde h_i(k)\;,\\
  \label{overlap2}
  \langle k|a_i\rangle        &= 
        \Theta(k-k_{i-1})\Theta(k_{i}-k) h_i(k)\;,
\end{align}
\end{subequations}
with $k_1\!<\!k_2\!<\cdots <\! k_N$.
Here $h_i(k)$ and $\tilde h_i(k)$ are functions on the domain 
$[k_{i-1},k_i)$.
The whole domain in $k$ is reconstructed with the union of
$N$ non-overlapping intervals, namely
\begin{equation}
  \label{krange}
  [0,\infty) = [k_0,k_1)\cup [k_1,k_2) \cdots \cup [k_{N-1},k_N)\;,
\end{equation}
where $k_0\!\equiv\!0$.
Being the free propagator $\hat G_0(z)$ diagonal in momentum space, 
$\hat G_0(z)\!=\!1/(z-\hat p^2)$, then
\begin{equation}
  \label{free}
  \langle \tilde a_i|\hat G_0(z)|a_j\rangle =  \delta_{ij}
  \langle \tilde a_i|\hat G_0(z)|a_i\rangle\;.
\end{equation}

Let us now consider the Lippmann-Schwinger 
integral equation for the scattering $\hat T$ matrix
\begin{equation}
\label{tu}
\hat T(z) = \hat V + \hat V\hat G_0(z) \hat T(z) \;.
\end{equation}
Replacing $\hat V$ from Eq.~\eqref{rank-n} and factorizing by $\hat T$
on the left we get
\begin{equation}
\label{tu2}
   \left [
  1 - \sum_i |\tilde a_i\rangle \lambda_i\langle a_i|\hat G_0(z)
   \right ] 
  \;
  \hat T(z) =  \sum_j |\tilde a_j\rangle \lambda_j\langle a_j| \;.
\end{equation}
Hence
\begin{equation}
\label{tu3}
  \hat T(z) =
  \left [
  1 - \sum_i |\tilde a_i\rangle \lambda_i\langle a_i| \hat G_0(z)
  \right ]^{-1} \sum_j |\tilde a_j\rangle \lambda_j\langle a_j| \;.
\end{equation}
Considering that
$\langle\tilde a_i|\hat G_0(z)|a_j\rangle\!=\!0$, when $i\!\neq\!j$,
then the above expression for $\hat T$ yields
\begin{equation}
\label{tu4}
  \hat T(z) = \sum_{i=1}^{N} \frac{|\tilde a_i\rangle \lambda_i\langle a_i|}
  {
  1 - \lambda_i\langle a_i| \hat G_0(z)|\tilde a_i\rangle 
}\;.
\end{equation}
Thus, the scattering matrix is also separable of rank-$N$.

When the above expression for $\hat T$ is projected in momentum
space as $\langle k|\cdots|k\rangle$, with $k$ in the $i$-th
interval, then
%restricted to $k\in[k_{i-1},k_i)$, then
\begin{equation}
  \label{tz}
  \langle k|\hat T(z)|k\rangle 
  \left [
    1 - \lambda_i \langle a_i| \hat G_0(z) | \tilde a_i\rangle
    \right ]
  = \lambda_i \langle k|\tilde a_i\rangle \langle a_i|k\rangle \;.
\end{equation}
Therefore the equations for the form factors on each of the
intervals get decoupled from one another.

%\nocite{*}
\section*{Acknowledgements}
The authors thank the 
partial support provided by the supercomputing infrastructure 
of the NLHPC (ECM-02): Powered@NLHPC.
N.A.A. acknowledges funding 
under doctoral fellowship ANID-Subdirecci\'on de Capital 
Humano/Doctorado Nacional/2019-21191174.
H.F.A. is grateful for the hospitality received at CEA-DAM,
Bruy\`eres-le-Ch\^atel, where part of this work was done.

%
% BibTeX users please use
%=================================
%\bibliographystyle{unsrt}
%\bibliography{misreferencias.bib}

%=================================
  \end{document}